\documentclass[useAMS,usenatbib]{mn2e}

\usepackage{amsmath}
\usepackage{graphicx} 
\usepackage{comment}

\newcommand{\Pii}{\mathrm{\pi}}
\newcommand{\ddp}[2]{\frac{\partial #1}{\partial #2}}
\newcommand{\lddp}[2]{\partial #1/\partial #2}

\newcommand{\eexp}{\mathrm{e}}

\newcommand{\sech}{\mathrm{sech}}

\newcommand{\Rb}{R_{\mathrm{b}}}
\newcommand{\Rd}{h_{\mathrm{R}}}
\newcommand{\Rs}{R_{\mathrm{s}}}
\newcommand{\Omegab}{\Omega_{\mathrm{b}}}

\newcommand{\Kpc}{~\mathrm{kpc}}
 
\newcommand{\Gyr}{~\mathrm{Gyr}}

\newcommand{\kmsec}{~\mathrm{km}~\mathrm{s}^{-1}}
\newcommand{\kmseckpc}{~\mathrm{km}~\mathrm{s}^{-1}~\mathrm{kpc}^{-1}}

\newcommand{\vc}{v_{\mathrm{c}}}

\newcommand{\de}{\mathrm{d}}

\newcommand{\RNum}[1]{\uppercase\expandafter{\romannumeral #1\relax}}

\newcommand{\ti}{t_\mathrm{i}}
\newcommand{\te}{t_\mathrm{e}}

\newcommand{\dd}[2]{\frac{\de #1}{\de #2}}
\newcommand{\pare}[1]{\left(#1\right)}
\newcommand{\paresq}[1]{\left[#1\right]}
\newcommand{\parec}[1]{\left\{#1\right\}}
\newcommand{\av}[1]{\langle #1 \rangle}

\newcommand{\img}{\mathrm{i}}
\newcommand{\Rep}{\operatorname{Re}}
\newcommand{\Eq}[1]{Eq.~(\ref{#1})}
\newcommand{\Eqs}[2]{Eqs.~(\ref{#1})-(\ref{#2})}

\newcommand{\Fig}[1]{Fig.~\ref{#1}}

\newcommand{\Phia}{\Phi^\mathrm{a}}

\newcommand{\dvz}{\Delta \av{v_z}}
\newcommand{\duz}{\Delta \av{u_z}}
\newcommand{\vzp}{\av{v_z}_\mathrm{p}}
\newcommand{\vzn}{\av{v_z}_\mathrm{n}}
\newcommand{\trho}{\tilde{\rho}}

\newcommand{\rhoa}{\tilde{\rho}^\mathrm{a}}
\newcommand{\URa}{U_R^\mathrm{a}}
\newcommand{\Uphia}{U_\phi^\mathrm{a}}
\newcommand{\uRa}{u_R^\mathrm{a}}
\newcommand{\uphia}{u_\phi^\mathrm{a}}
\newcommand{\Sec}[1]{Section~\ref{#1}}
\newcommand{\App}[1]{Appendix~\ref{#1}}
\newcommand{\Or}{\mathrm{O}}

\newcommand{\uro}{u_{R,1}}
\newcommand{\uphiz}{u_{\phi,0}}
\newcommand{\uphio}{u_{\phi,1}}
\newcommand{\uzo}{u_{z,1}}

\newcommand{\URo}{U_{R,1}}
\newcommand{\Uphio}{U_{\phi,1}}
\newcommand{\Uphiz}{U_{\phi,0}}

\newcommand{\calF}{{\cal F}}
\newcommand{\calG}{{\cal G}}
\newcommand{\tOm}{\tilde{\Omega}}
\newcommand{\tkappa}{\tilde{\kappa}}
\newcommand{\tDelta}{\tilde{\Delta}}
\newcommand{\tB}{\tilde{B}}
\newcommand{\Li}{\mathrm{Li}}

\title[The vertical effects of disc non-axisymmetries from
  perturbation theory]{The vertical effects of disc non-axisymmetries
  from perturbation theory: the case of the Galactic bar}
\author[G. Monari et al.]  {
  Giacomo~Monari\thanks{Email:~\texttt{giacomo.monari@astro.unistra.fr}},
  Benoit~Famaey, Arnaud~Siebert\\
  Observatoire astronomique de
  Strasbourg, Universit\'e de Strasbourg, CNRS UMR 7550, 11 rue de
  l'Universit\'e, 67000 Strasbourg, France } \date{Released 2015 Xxxxx
  XX}

\pagerange{\pageref{firstpage}--\pageref{lastpage}} \pubyear{2015}

\def\LaTeX{L\kern-.36em\raise.3ex\hbox{a}\kern-.15em
    T\kern-.1667em\lower.7ex\hbox{E}\kern-.125emX}

\begin{document}

\label{firstpage}

\maketitle

\begin{abstract}
Evidence for non-zero mean stellar velocities in the direction
perpendicular to the Galactic plane has been accumulating from various
recent large spectroscopic surveys. Previous analytical and numerical
work has shown that a ``breathing mode" of the Galactic disc, similar
to what is observed in the Solar vicinity, can be the natural
consequence of a non-axisymmetric internal perturbation of the
disc. Here we provide a general analytical framework, in the context
of perturbation theory, allowing us to compute the vertical bulk
motions generated by a single internal perturber (bar or spiral
pattern). In the case of the Galactic bar, we show that these
analytically predicted bulk motions are well in line with the outcome
of a numerical simulation. The mean vertical motions induced by the
Milky Way bar are small (mean velocity of less than $1\kmsec$) and cannot
be responsible alone for the observed breathing mode, but they are
existing. Our analytical treatment is valid close to the plane for all
the non-axisymmetric perturbations of the disc that can be described
by small-amplitude Fourier modes. Further work should study how the
coupling of multiple internal perturbers and external perturbers is
affecting the present analytical results.
\end{abstract}


\section{Introduction}

The kinematics of stars perpendicular to the Galactic plane
traditionally had (and still has) great importance, for instance as a
probe of the vertical force and mass density in the Solar neighborhood
(e.g.,
\citealt{KuijkenGil91,Creze98,Siebert03,Bienayme14,Read2014}). In
dynamical modeling, the natural zeroth order approximation is to
assume a null mean vertical motion everywhere in the Galaxy (as it
should be in a steady-state axisymmetric stellar system, see
\citealt{BT2008}). However, thanks to the spatial extension and
accuracy of recent surveys, it has been discovered that significant
non-zero mean vertical motions do exist in the extended Solar
neighborhood (\citealt{Widrow2012,Williams2013,Carlin2013}).  These
consist in patterns looking like ``rarefaction-compression'' waves or
``breathing modes" of the disc. Whilst originally associated solely
with excitations by external sources such as a passing satellite
galaxy or a small dark matter substructure crossing the Galactic disc
(\citealt{Widrow2012}, \citealt{Gomez2013,YannyGardner2013,Feldmann}),
it was then shown that if the density perturbation has even parity
with respect to the Galactic plane (i.e., is plane-symmetric), and the
vertical velocity field has odd parity (i.e., a breathing mode). Then
{\it internal} non-axisymmetric perturbations, such as spiral arms,
could be sufficient to cause this (\citealt{Faure2014}, hereafter
F14). This was shown analytically by solving the linearized Euler
equations in a cold fluid toy model, and confirmed through
test-particle simulations for quasi-static spiral arms. Subsequently,
the same effect was found in self-consistent simulations
\citep{Debattista2014}. Such an effect from internal non-axisymmetries
is of course not mutually incompatible with external perturbers
playing a role in shaping the velocity field \citep[e.g.][in which
  satellite interactions cause both bending and breathing modes
  depending on the velocity of the satellite]{Widrow2014}, and such
perturbers are themselves exciting non-axisymmetric modes such as
spiral waves \citep{Widrow2015}.

There is a long history of theoretical studies of the dynamical
effects of disc non-axisymmetries in two dimensions in the Galactic
plane, dating back to the seminal works of, e.g., \cite{LinShu} and
\cite{Toomre} on spiral arms. From the observational point of view,
striking kinematic features related to the bar and spiral arms were
for instance found in the Solar neighbourhood in the form of moving
groups, i.e. local velocity-space substructures made of stars of very
different ages and chemical compositions
\citep{Dehnen,Chereul,Famaey,Antoja}. At a less fine-grained level, it
is obvious that such local velocity-space substructures will affect
the {\it mean} motions too. And indeed, it has long been well
established that a non-zero mean radial motion is a natural response
to disc instabilities such as spiral arms (e.g., \citealt{LinShu}, see
also \citealt{BT2008}), independently of the exact nature of the
perturber (quasi-static or transient). The same is true for the
effects of the Galactic bar across the whole Galactic disc (see, e.g.,
\citealt{Kuijken1991}).

Such non-zero mean radial motions have recently been detected in the
extended Solar neigbourhood with the RAVE survey \citep{Siebert2011},
and shown to be consistent with the effect of spiral arms in the form
of Lin-Shu density waves \citep{Siebert2012}. Velocity fluctuations
have also been detected on larger scales with the APOGEE survey, and
attributed to the effects of the Galactic bar
\citep{Bovy2015}. \citet[][hereafter M14]{Monari2014}, found that the
bar can also explain the RAVE radial velocity gradient of
\cite{Siebert2011}, but did not find hints of the observed vertical
bulk motions \citep{Williams2013}. On the contrary, the spiral arm
model of F14 has been shown to generate a ``breathing mode''
qualitatively similar to what is observed for vertical
motions. Generally speaking, theoretical studies of the dynamical
response to disc non-axisymmetries in 3D are much less developed than
in 2D (see however \citealt{Patsis}; \citealt{Cox}).

In this paper, our aim is to provide a general analytical framework
linking the vertical bulk motions generated by disc non-axisymmetries
to the horizontal bulk motions. This analytic treatment goes beyond
the analytic toy-model of F14 for a pressureless three-dimensional fluid, which
was only suitable for an extremely cold stellar population, and which
required a few unrealistic assumptions (for instance that the vertical
force of the axisymmetric potential was negligible w.r.t. to the
maximum vertical force of the perturber). Our reasoning hereafter is
based on a linearization of the zeroth order moment of the
collisionless Boltzmann equation, i.e., the continuity equation for
stellar systems. As an improvement over the analytic treatment in F14,
the present calculation is valid for the whole range of velocity
dispersions compatible with the epicyclic approximation, and it is
valid close to the plane for all non-axisymmetric perturbations of the
potential described by small-amplitude Fourier modes. In particular,
we show that even the Galactic bar is expected to induce vertical bulk
motions in the Galactic disk. We use this particular case to compare
our analytic results with a numerical study.

The paper is organized as follows. In \Sec{sect:dynamics} we linearize
the continuity equation, deriving a theoretical prediction for the
vertical bulk motion. More specifically, we relate it to the
horizontal motions in the case where the potential perturbation is a
Fourier mode of small amplitude. In \Sec{sect:sim} we present the
outcome of a numerical test-particle simulation with a realistic model
of the Galactic disc under the influence of the gravity of the Milky
Way and the bar, and compare it with the analytical results. In
\Sec{sect:concl} we conclude.

\section{Vertical effects of perturbations}\label{sect:dynamics}

\subsection{The linearized continuity equation}

Using the cylindrical coordinates $\pare{R,\phi,z}$ and velocities
$\pare{v_R,v_\phi,v_z}=\pare{\dot{R},R\dot{\phi},\dot{z}}$, and
integrating the collisionless Boltzmann equation over velocity space,
one gets the following form of the continuity equation for stellar
systems:
\begin{equation}\label{eq:continuity}
  \ddp{\rho}{t}+\frac{1}{R}
  \ddp{\pare{R\rho u_R}}{R}+
  \frac{1}{R}\ddp{\pare{\rho u_\phi}}{\phi}+
  \ddp{\pare{\rho u_z}}{z}=0,
\end{equation}
where $\rho$ is the density of the system in configuration space, and
$u_i\equiv\av{v_i}$ are the average $v_i$ velocities, all estimated at
$\pare{R,\phi,z}$ and at the time $t$.

For an axisymmetric stationary system, symmetric with respect to
the plane $z=0$, all observables depend on $\pare{R,z}$ only, and one has
\begin{equation}
  \rho=\rho_0,\quad u_R=0,\quad u_\phi=\uphiz,\quad u_z=0.
\end{equation}

Let us now consider the case where the system is perturbed by a small
non-axisymmetric perturbation in the potential, i.e.,
\begin{equation}\label{pottot}
  \Phi\pare{R,\phi,z,t}=\Phi_0\pare{R,z}+\epsilon\Phi_1\pare{R,\phi,z,t},
\end{equation}
where $\Phi_0$ is the unperturbed axisymmetric potential, $\epsilon
\ll 1$, and $\Phi_1$ has the same order of magnitude as $\Phi_0$.

Considering a small response to this small perturbation, we can write:
\begin{equation}\label{eq:pert}
  u_R=\epsilon \uro,\quad u_\phi=\uphiz+\epsilon
  \uphio,\quad u_z=\epsilon \uzo,\quad \rho=\rho_0+\epsilon \rho_1.
\end{equation}
Plugging \Eq{eq:pert} into \Eq{eq:continuity} and dropping the terms
that are $\Or\pare{\epsilon^2}$, we obtain the well-known linearized
continuity equation:
\begin{equation}\label{eq:continuity_2}
  \ddp{\rho_1}{t}+\frac{1}{R}
  \ddp{\pare{R\rho_0 \uro}}{R}+
  \frac{\rho_0}{R}\ddp{\uphio}{\phi}+
  \frac{\uphiz}{R}\ddp{\rho_1}{\phi}=
  -\ddp{\pare{\rho_0 \uzo}}{z}.
\end{equation}
From this equation, it is immediately apparent that, for a given
background axisymmetric model, one will be able to relate the vertical
bulk motion $\uzo$ to the horizontal responses $\uro$ and $\uphio$ and
the density wake of the perturbation $\rho_1$.

\subsection{Solution for an exponential disc}

Disc non-axisymmetries respect the plane symmetry, hence on the
galactic plane $\uzo(z=0)=0$. This would not be true in the case where
a bending mode, due to satellite interactions for example, is
present \citep{Xu, Widrow2015}. This will be the subject of further
work.

From $\uzo(z=0)=0$, the solution of \Eq{eq:continuity_2} reads
\begin{equation}\label{eq:rhouz1}
  \rho_0\pare{R,z}\uzo\pare{R,\phi,z,t}=-\int_0^{z}\calF\pare{R,\phi,\xi,t}\de\xi,
\end{equation}
where 
\begin{equation}\label{eq:fA}
  \calF\pare{R,\phi,z,t}\equiv\ddp{\rho_1}{t}+
  \frac{1}{R}\ddp{\pare{R\rho_0 \uro}}{R}+
  \frac{\rho_0}{R}\ddp{\uphio}{\phi}+
  \frac{\uphiz}{R}\ddp{\rho_1}{\phi}.
\end{equation}
We now assume that the unperturbed stellar system that we consider is an
axisymmetric exponential disc, i.e.,
\begin{equation}
  \rho_0\pare{R,z}=\rho_0\pare{0,0}\exp\pare{-\frac{R}{\Rd}-\frac{|z|}{h_z}}
\end{equation}
where $\Rd$ and $h_z$ are the scale length and height respectively. We
also assume that $\rho_1$ and $\rho_0$ have the same exponential
vertical dependence, i.e., $\trho\equiv \rho_1/\rho_0$ is constant
with $z$.
With these assumptions \Eq{eq:rhouz1} becomes
\begin{equation}\label{eq:rhouz2}
  \uzo\pare{R,\phi,z,t}=-\frac{\int_0^{z}\eexp^{-|\xi|/h_z}\calG\pare{R,\phi,\xi,t}\de\xi}{\eexp^{-|z|/h_z}},
\end{equation}
where
\begin{equation}\label{eq:G}
  \calG\pare{R,\phi,z,t}\equiv\ddp{\trho}{t}+
  \frac{\uphiz}{R}\ddp{\trho}{\phi}
  +\frac{\uro}{R}-\frac{\uro}{\Rd}+\ddp{\uro}{R}+\frac{1}{R}\ddp{\uphio}{\phi}.
\end{equation}

\subsection{Taylor expansion close to the plane}
We now indicate only the dependence on $z$ of the functions to
simplify the notation, and Taylor expand $\calG$ in powers of $z$
up to second order close to the plane, i.e., from \Eq{eq:rhouz2},
\begin{equation}\label{eq:rhouz3}
  \uzo\pare{z}\approx-\frac{\int_0^{z}\eexp^{-|\xi|/h_z}
  \paresq{\calG\pare{0}+\frac{1}{2}\ddp{^2\calG}{z^2}(0)\xi^2}\de\xi}{\eexp^{-|z|/h_z}}.
\end{equation}
Therefore, from \Eq{eq:pert} and \Eq{eq:rhouz3},
\begin{align}\label{eq:uz}
  u_z(z) &=\epsilon~\mathrm{sgn}(z)h_z\biggl[
    \calG(0)\pare{1-\eexp^{|z|/h_z}} \nonumber \displaybreak[0] \\
    &+\ddp{^2\calG}{z^2}(0)
    \pare{\frac{z^2}{2}+h_z|z|+h_z^2-\eexp^{|z|/h_z}h_z^2}\biggr],
\end{align}
and, averaging $u_z$ over $z>0$, weighted by $\rho$, we obtain
\begin{align}\label{eq:uzint}
  &\duz \pare{z}\equiv 2\frac{\int_0^z\rho\pare{\xi} u_z\pare{\xi}~\de
    \xi} {\int_0^z \rho\pare{\xi} ~\de \xi}= \nonumber \displaybreak[0] \\ 
  &=2\epsilon\Biggl\{\paresq{\calG(0)(h_z-z)+\ddp{^2\calG}{z^2}(0)h_z^2\pare{3h_z-z}} \nonumber \displaybreak[0] \\
  &-z\paresq{\calG(0)+
        \ddp{^2\calG}{z^2}(0)\frac{h_z}{2}\pare{6h_z+z}}/\pare{\eexp^{z/h_z}-1}\Biggr\},
\end{align}
the difference between the mean vertical velocities of stars within
$z$, above and below the Galactic plane, since $u_z(z)=-u_z(-z)$. A
value $\duz>0$ ($\duz<0$) implies that the stars of both Galactic
hemispheres tend to move away from (towards) the Galactic plane, while
for $\duz=0$ we have $u_z=0$, like in an axisymmetric Galaxy.

Note that, whilst being $C^0$ continuous, the function $u_z(z)$ given
by \Eq{eq:uz} is not continuously differentiable ($C^1$) at
$z=0$. This is a consequence of the fact that the vertical density
distribution $\exp\pare{-|z|/h_z}$ is itself not $C^1$ at $z=0$. This
technical issue disappears for a $\sech^2(z/h_z)$ density
distribution, which is $C^1$ at $z=0$. In \App{app:sech} we compute
the formulae corresponding to \Eqs{eq:uz}{eq:uzint} in the $\sech^2$
case, and show that the results are quantitatively similar for the
same value of $h_z$ close to the plane.

\subsection{Computing the horizontal bulk motions}

At this point, we will need to estimate $\calG(0)$ and
$\lddp{^2\calG(0)}{z^2}$. To do so, we consider the potential
perturbation $\Phi_1$, the density response $\trho$, and the
horizontal mean velocities as Fourier $m$-modes propagating in the disc, i.e.,
\begin{equation}\label{eq:phi1fourier}
  \Phi_1\pare{R,\phi,z,t}=\Rep\parec{\Phia(R,z)\exp\paresq{\img m\pare{\phi-\Omegab t}}}.
\end{equation}
\begin{equation}\label{eq:rho}
  \trho\pare{R,\phi,t}=\Rep\parec{\rhoa(R)\exp\paresq{\img m\pare{\phi-\Omegab t}}},
\end{equation}
\begin{equation}\label{eq:uro}
  \uro\pare{R,\phi,z,t}=\Rep\parec{\uRa(R,z)\exp\paresq{\img m\pare{\phi-\Omegab t}}},
\end{equation}
\begin{equation}\label{eq:uphi}
  \uphio\pare{R,\phi,z,t}=\Rep\parec{\uphia(R,z)\exp\paresq{\img m\pare{\phi-\Omegab t}}},
\end{equation}
where we have reintroduced the $R$, $\phi$, and $t$ explicit
dependencies. Dividing \Eq{eq:continuity_2} by $\rho_0$ and integrating
over $z$ we obtain the 2D continuity equation,
i.e.,
\begin{equation}\label{eq:2Dcont}
  \ddp{\trho}{t}+
  \frac{\Uphiz}{R}\ddp{\trho}{\phi}+
  \frac{\URo}{R}-\frac{\URo}{h_R}+\ddp{\URo}{R}+\frac{1}{R}\ddp{\Uphio}{\phi}=0,
\end{equation}
where
\begin{equation}
  U_i\equiv \frac{\int_{-\infty}^\infty\rho\pare{z} u_i\pare{z}~\de
    \xi}{\int_{-\infty}^\infty\rho\pare{z}~\de z},
\end{equation}
and $u_i$ is one of the $\uro$, $\uphio$, $\uphiz$, $\uRa$, and
$\uphia$, functions defined above. Using the condition \Eq{eq:2Dcont}
we can derive $\rhoa$ as
\begin{equation}\label{eq:rhoa}
  \rhoa(R)=\frac{m\Uphia/R+\img\pare{\URa/h_R-\URa/R-\lddp{\URa}{R}}}
       {m\pare{\Omegab-\Uphiz/R}}. 
\end{equation}

The perturbative regime cannot give accurate estimates of $\rho\uRa$,
$\rho\uphia$, and $\rho\uphiz$ far from the $z=0$ plane. However,
typically these functions must decrease quite fast as a function of
$z$ and tend to $0$ as $z$ goes to infinity. Therefore, we can
approximate
\begin{equation}\label{eq:av}
  U_i\pare{R,\phi,t}\approx\frac{\int_{-\zeta}^\zeta\rho(R,\xi)u_i\pare{R,\phi,\xi,t}~\de\xi}
  {\int_{-\zeta}^\zeta\rho(R,\xi)~\de\xi},
\end{equation}
where $\zeta$ is the height from the plane up to which we are
computing the average and at which the integral in the perturbative
regime is converging to a constant value\footnote{In the rest of the
  paper we adopt $\zeta=1\Kpc$, as it is sufficient for the
  convergence of the $\rho u_i$ functions in the barred case of
  \Sec{sect:sim}.}.

We then obtain the value of $\uRa$ and $\uphia$ close the $z=0$ plane by
linearizing Jeans equations, in the same way than
\cite{Kuijken1991} in the 2D case. Under the assumptions that the
velocity dispersions $\sigma_{R}^2$ and $\sigma_\phi^2$ are related by
the epicyclic approximation (\citealt{BT2008}), that the radial scale
length of $\Phi_1$ is larger than
$\max\paresq{|\uro|,\sigma_R}/\kappa$, and that the mixed terms in the
velocity dispersion $\sigma_{R,z}$ and $\sigma_{\phi,z}$ are
negligible, one obtains the two following equations for $\uRa$ and $\uphia$:
\begin{subequations}
  \begin{equation}\label{eq:uRa}
    \uRa\pare{R,z}=\frac{\img}{\tDelta}
    \paresq{m\pare{\Omegab-\tOm}\ddp{\Phia}{R}-\frac{2m\tOm}{R}\Phia},
  \end{equation}
  \begin{equation}\label{eq:uphia}
    \uphia\pare{R,z}=-\frac{1}{\tDelta}
    \paresq{2\tB\ddp{\Phia}{R}+\frac{m\pare{m\Omegab-m\tOm}}{R}\Phia},
  \end{equation}
\end{subequations}
where
\begin{subequations}\label{eq:quant}
  \begin{equation}
    \tOm(R,z)=\frac{\uphiz(R,z)}{R},
  \end{equation}
  \begin{equation}
    \tkappa(R,z)^2=R\ddp{\tOm^2}{R}(R,z)+4\tOm^2(R,z),
  \end{equation}
  \begin{equation}
    \tDelta(R,z)=\tkappa^2(R,z)-m^2\paresq{\Omegab-\tOm(R,z)}^2,
  \end{equation}
  \begin{equation}
    \tB(R,z)=-\tOm(R,z)-\frac{1}{2}R\ddp{\tOm}{R}(R,z).
  \end{equation}
\end{subequations}
Notice how the quantities \Eq{eq:quant} would yield the well known
$\Omega$, $\kappa$, $\Delta$, $B$ functions (\citealt{BT2008}) on the
$z=0$ plane for a zero-asymmetric-drift cold fluid.

The assumption that the radial scale length of $\Phi_1$ is larger than
$\max\paresq{|\uro|,\sigma_R}/\kappa$ is in general not valid for
spiral arms, except for very cold populations. Nevertheless, our
framework for relating the vertical and horizontal motions remains
very general. In the case of spirals we should use the generalized
equations for $\uRa$ and $\uphia$ from \cite{LinShu}, including
reduction factors depending on the velocity dispersion of the
background stellar population \citep{BT2008}. Hereafter, we will
concentrate on the case of the bar, where the above assumption on the
radial variation of $\Phi_1$ remains valid even for a realistic
background stellar velocity dispersion.

\subsection{Relating the vertical and horizontal bulk motions}

To get the vertical bulk motions, we need $\calG(0)$ and
$\lddp{\calG(0)}{z^2}$ in order to evaluate $\duz(z)$ in
\Eq{eq:uzint}. To do so, we need $\rhoa$ and its derivatives.  Hence,
we further simplify the integrals in \Eq{eq:av} by expanding, up to
the second order, the functions $u_i$ in \Eq{eq:uRa} and
\Eq{eq:uphia}. These are even function with the respect of $z$, so
\begin{equation}
  u_i(R,z)\approx u_i(R,0)+\frac{1}{2}\ddp{^2u_i}{z^2}(R,0)z^2.
\end{equation}
For the specific function $\uphiz$ coming from the background
axisymmetric model, in \App{app:uphiz} we estimate
\begin{equation}
  \ddp{^2\uphiz}{z^2}(R,0)\approx \frac{R}{\uphiz(R,0)}\dd{\nu^2}{R}(R),
\end{equation}
where $\nu(R)\equiv \sqrt{\lddp{^2\Phi_0(R,0)}{z^2}}$ is the vertical
epicyclic frequency.

We thus now have a fully analytical expression for $\duz(z)$ in
\Eq{eq:uzint}, which we can apply to the case of the Galactic bar
hereafter. Indeed, according to the above calculations, non-zero
vertical motions should be induced by {\it any} non-axisymmetric
perturbations described by small-amplitude Fourier modes, even though
M14 had not found them in a bar simulation. This means that they are
probably quite small, but they should be present: here, we can
explicitly estimate them analytically, and then check for their
presence in a numerical simulation similar to that of M14. Note
however that the present analytical perturbative approach breaks down
close to the corotation and Lindblad resonances where our analytical
expression for $\duz$ diverges. Note also that we consider here only
the response to the potential perturbation, and do not relate back the
density response to the potential perturbation.

\section{Application to the Galactic bar}\label{sect:sim}

\subsection{Galactic potential and disc stellar population}\label{sect:pot}
We take for the background axisymmetric potential $\Phi_0$ the Milky
Way Model~I by \cite{BT2008}.  It consists of a spheroidal dark halo
and bulge, and three disc components: thin, thick, and ISM disc. The
disc densities decrease exponentially with Galactocentric radius and
height from the Galactic plane. The position of the Sun in this model
is $(R_0,z_0)=(8\Kpc,0)$. In this axisymmetric potential, we consider
a background stellar population described by an exponential disc of
scale length $\Rd=2\Kpc$ and scale height $h_z=0.3\Kpc$, with radial and vertical velocity
dispersions varying as $\sigma_R =
\sigma_{R,0}\exp\paresq{-\pare{R-R_0}/\Rs}$, $\sigma_z =
\sigma_{z,0}\exp\paresq{-\pare{R-R_0}/\Rs}$, where
$(\sigma_{R,0},\sigma_{z,0})=(35,15)\kmsec$, and $\Rs=5\Rd$. The mean
tangential velocity has an asymmetric drift described by Stromgren's
formula (\citealt{BT2008}) adapted to our case, i.e.,
\begin{equation}
  \uphiz(R,0)=\vc-\frac{\sigma_R^2}{2\vc}
  \pare{\frac{\kappa^2}{4\Omega^2}-1+\frac{7}{5}\frac{R}{\Rd}}.
\end{equation}

\subsection{Bar potential}
We describe the perturbation due to the bar by a 3D extension of the
quadrupole bar used by \cite{Weinberg1994} and \cite{Dehnen2000},
i.e. $\Phi_1$ as described in \Eq{eq:phi1fourier} with $m=2$ and the
angle $\phi$ measured from the long axis of the bar, with
\begin{equation}\label{phia}
  \Phia(R,z) = \frac{V_0^2}{3} \left( \frac{R_0}{\Rb} \right)^3 \frac{R^2}{r^2} {\cal U}(r)
\end{equation}
and
\begin{equation}
  {\cal U}(r) = \left\{
  \begin{array}{l l}
    -(r/\Rb)^{-3} & \quad \text{for } r \geq \Rb,\\
    (r/\Rb)^{3}-2 & \quad \text{for } r < \Rb,
  \end{array} \right.
\end{equation}
where $R_0$ is the Galactocentric radius of the Sun, $V_0=\vc(R_0)$
the local circular velocity, $R_b$ the bar length, and $r^2=R^2+z^2$.

The amplitude $\epsilon$ in \Eq{pottot} then represents the ratio
between the bar and axisymmetric background radial forces at
$\pare{R,\phi,z}=\pare{R_0,0,0}$.

Here we choose $\epsilon=0.01$ (\citealt{Dehnen2000}),
$\Omegab=52.23\kmseckpc$, $\Omegab/\Omega(R_0)=1.89$,
(\citealt{Antoja2014}), and $\Rb=3.5\Kpc$ (\citealt{Dwek1995}).

\subsection{Analytical results}

\begin{figure}
  \centering
  \includegraphics[width=\columnwidth]{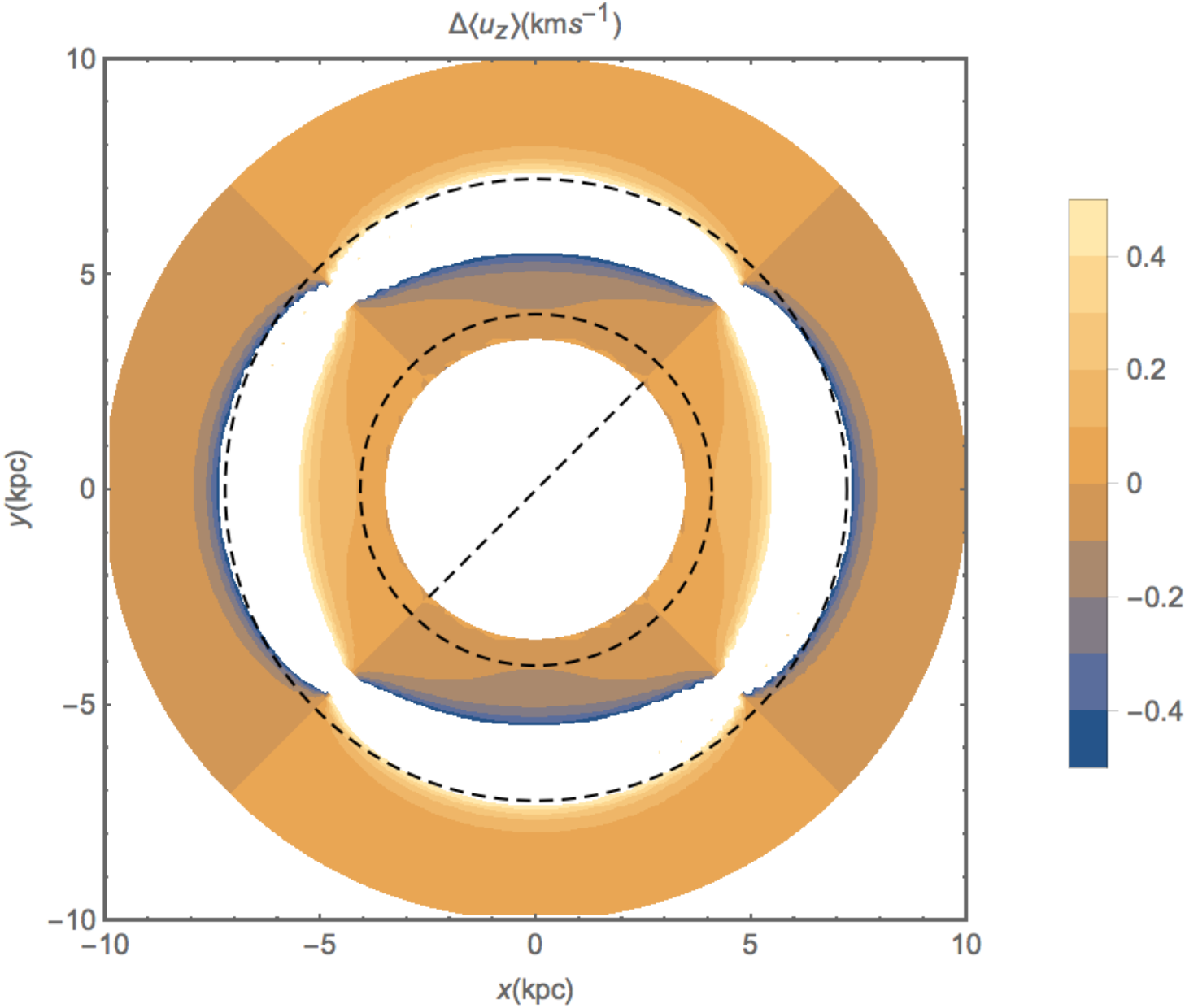}
  \caption{Difference between the average vertical motion of stars in
    the northern and southern hemisphere of the Milky Way $\duz$
    in our analytical model for $z=0.3\Kpc$, as a function of the $x$
    and $y$ positions in the plane. The dashed line corresponds to the
    long axis of the bar, and the dashed circles to the position of the
    corotation and OLR.}
  \label{fig:diffvz_th}
\end{figure}

With all these inputs, we can then immediately compute $\duz$ from
\Eq{eq:uzint}, and in \Fig{fig:diffvz_th} we plot the predicted
difference between the mean vertical velocities within 300~pc above
and below the Galactic plane, i.e. $\duz(R,\phi,z=0.3\Kpc,t=0)$.
The dashed line represents the orientation of the long axis of the
bar, and the dashed circles the position of the corotation and outer
Lindblad resonance (OLR), i.e., $R$ where $\Omega(R)=\Omegab$,
and $\paresq{\Omega(R)-\Omegab}+\kappa(R)/2=0$.

We are interested in studying the response of the background
axisymmetric thin disc to the bar potential, hence we focus on the
outside of the bar region ($R>\Rb$). We also limit ourselves to
$R<10\Kpc$ in this plot, since that is the interesting region where
there are still non-negligible effects. In this plot the Galaxy and
the bar rotate counter-clockwise.

\Fig{fig:diffvz_th} shows that the mean vertical motions induced by
the Milky Way bar are small, of the order of a few tenth of $\kmsec$,
and up to about $0.5\kmsec$. More importantly, it shows how these
motions depend on the angle from the long axis of the bar and the
distance from the center of the Galaxy. The angular dependence of
$\duz$ is that of an $m=2$ mode propagating in the disk while
keeping the same configuration w.r.t. the bar as the bar rotates.

As noted earlier, close to the corotation and OLR, $\duz$
diverges, because the linear theory presents poles there, as is
evident in \Eq{eq:rhoa}, and \Eqs{eq:uRa}{eq:uphia}. Higher order
expansions (and changing variables so that the new variables only vary
slowly there) would be required near these resonances, where we leave
our predictions as blank. Notice how the correspondence in the figure
to the resonances is not exact. This is due to the fact that, since we
consider a stellar disk rather than a cold fluid, in the analytical
model we use $\uphiz$ which differs from $\vc$ for the asymmetric
drift and the dependence from $z$, and shift the position of the
resonances in the model. At the OLR there is a
phase shift of $\Pii/2$ in $\duz$: along a given fixed azimuth
$\phi$, the value of $\duz$ inside and outside the OLR has
always opposite signs.  

At a given radius, the function $\duz$ also changes sign between the
regions ahead of the bar ($\phi>0$) and behind the bar
($\phi<0$). This is due to the fact that $u_z$ is an odd function in
$\phi$, as it is the sum of odd functions, as we can find out by
looking at Eqs.~(\ref{eq:G}), (\ref{eq:uzint}),
(\ref{eq:rho}-\ref{eq:uphi}), (\ref{eq:rhoa}), (\ref{eq:uRa}), and
(\ref{eq:uphia}). In particular, the compression ($\duz<0$) is ahead
of the bar between the corotation and the OLR, while it is behind the
bar outside the OLR, due to the change of sign of $\tDelta$ in
\Eq{eq:quant}.

We note that the pattern in $\duz$ is actually qualitatively following
the $\uro$ pattern, because they are both odd functions in $\phi$, as
can be seen from \Eq{eq:uro} and \Eq{eq:uRa} for $\uro$. This is in
stark contrast with what happens in the case of spiral arms, for which
F14 showed a clear phase shift between the vertical and radial bulk
motions. The reason for this difference is that $\Phia(R,z)$ is a pure
real function in the case of the bar, but depends on ${\rm exp}(i \, m
{\rm ln}(R)/{\rm tan} p)$ in the case of spirals\footnote{Where $p$ is
  the pitch angle}, inducing a phase shift in \Eq{eq:uro} and
\Eq{eq:uRa} for spirals.

\begin{figure}
  \centering
  \includegraphics[width=\columnwidth]{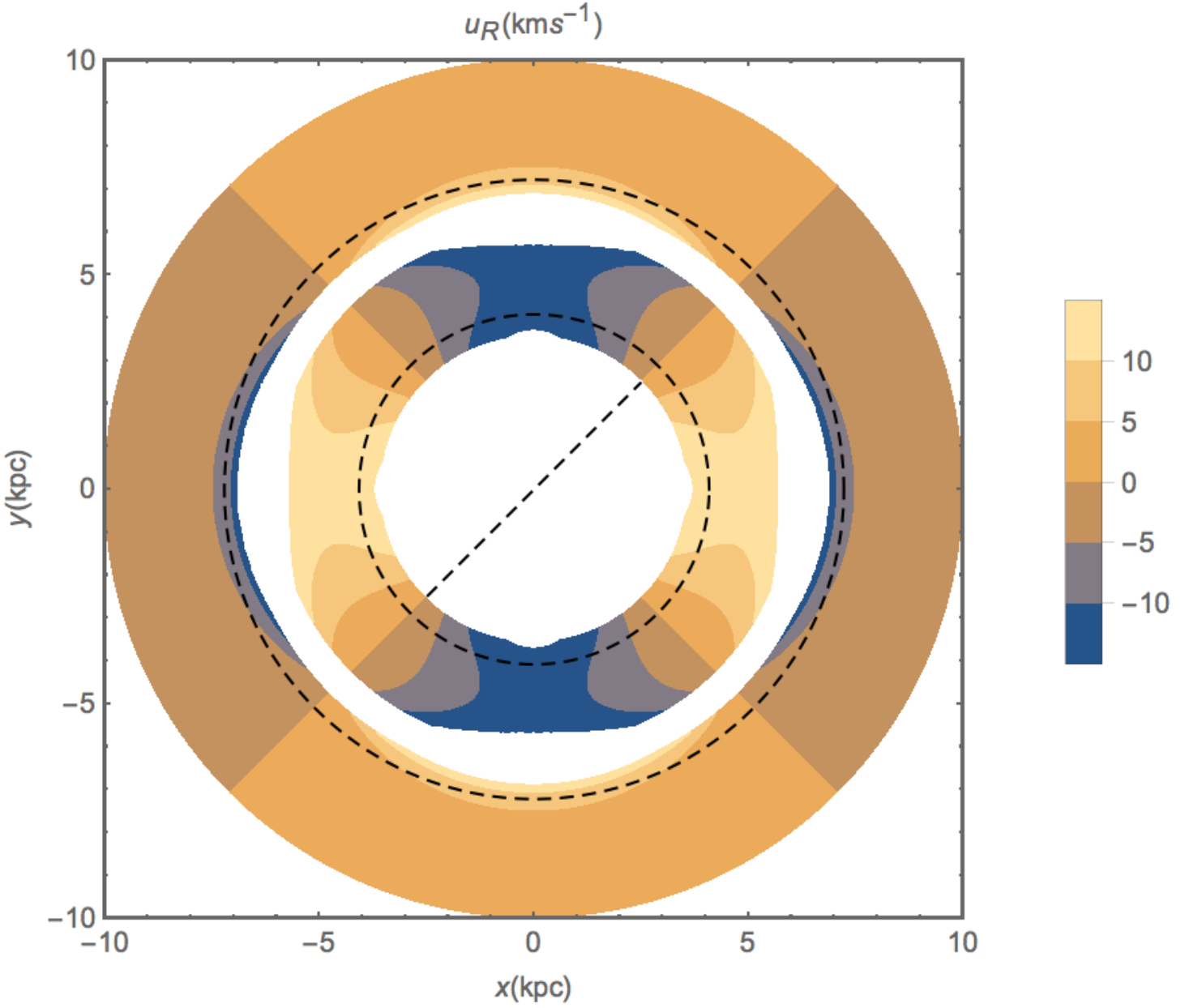}
  \includegraphics[width=\columnwidth]{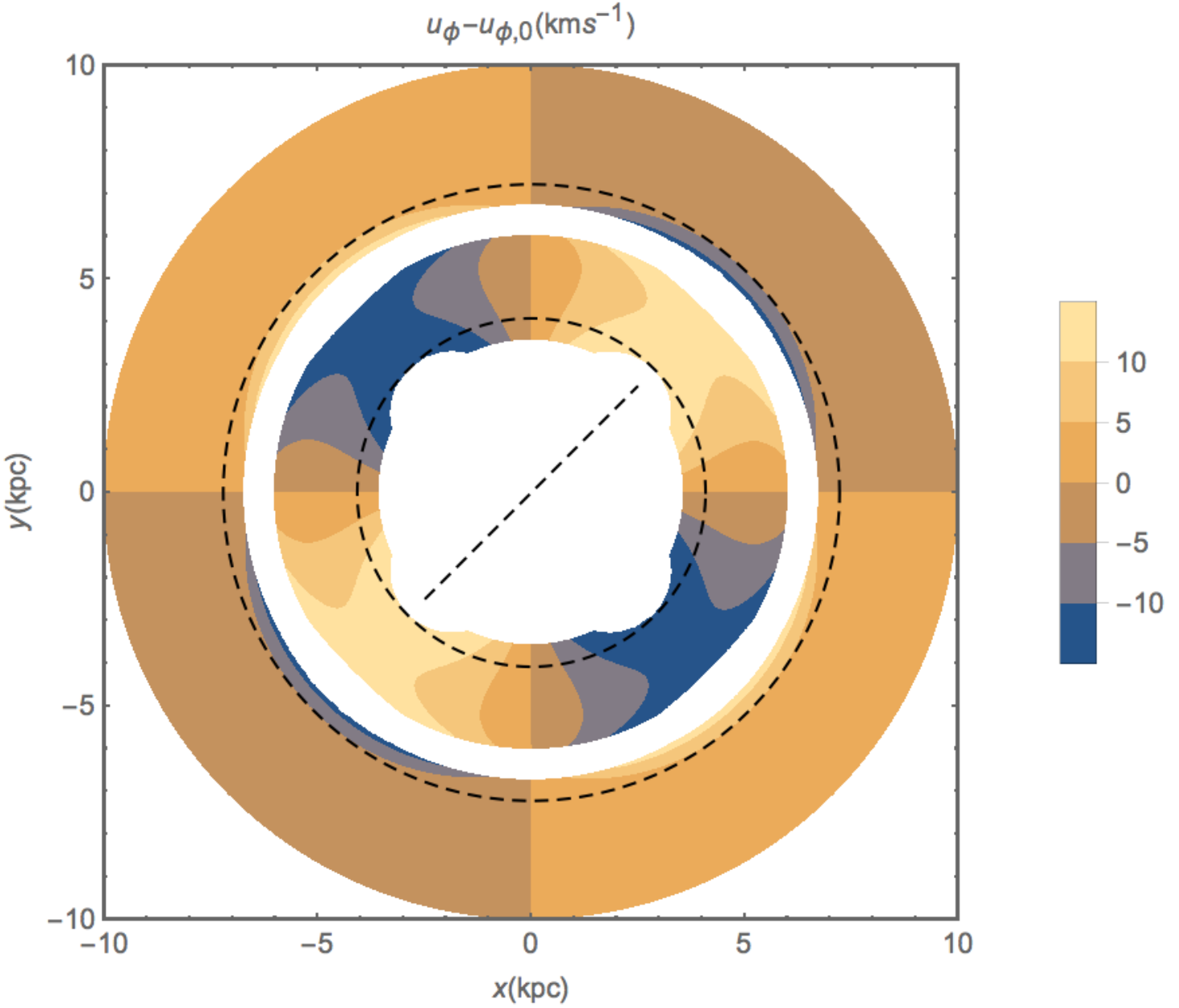}
  \caption{As in \Fig{fig:diffvz_th} for the horizontal motions $u_R$
    (top panel) and $u_\phi-\uphiz$ (bottom panel) in our analytical
    models for $z=0$.}
  \label{fig:hori-ana}
\end{figure}
For visual comparison with the $\duz$ pattern of \Fig{fig:diffvz_th},
we plot the analytical $u_R$ and $\epsilon \uphio$ patterns for the
bar model in \Fig{fig:hori-ana}. In particular, we see that between
the corotation and OLR and for $\phi<0$ there is a net outward motion
of stars ($u_R>0$) with vertical expansion, while for $\phi>0$ the
radial motion is inward ($u_R<0$) and there is a vertical
compression. The same behaviour has been seen in Fig.~12 of
\citet{Widrow2014}, where a satellite-induced bar has been shown to
create the same qualitative pattern. Outside the OLR, the pattern is
reversed, both in $u_R$ and $\duz$.

A heuristic explanation of this phenomenon is that the net radial
flows are related to the shape of the orbits induced by the bar inside
and outside the resonances (see \citealt{BT2008}). Outside the OLR,
the orbits in the bar frame are aligned with the bar and retrograde,
while they are perpendicular to the bar and retrograde between the
corotation and OLR: this naturally explains the shape of the radial
flow. Stars are then also adapting to the vertical restoring force,
stronger in the inner parts than the outer parts of the Galaxy: when
the flow is inward, the force is stronger and the vertical motion
corresponds to a compression, and when the flow is outward it
corresponds to an expansion-rarefaction. We note that the situation is
different in the case of spirals because of the restoring force from
the spiral potential itself.

Our analytical predictions for the vertical bulk motions induced by
the bar can now be compared to a simulation akin to M14 in the next
subsection.

\subsection{Numerical results}
\subsubsection{Initial conditions and integration time}
A simple and yet very effective way to study numerically the response
of a stellar disc to an external perturbation is through test-particle
simulations. These consist of the numerical integration of the
equations of motion of massless particles (representing the stars in
the disc), accelerated by a gravitational field (representing the
potential of the galaxy) which is uninfluenced by the particles
themselves.

We generate $5\times 10^7$ initial positions and velocities for our
simulations, as discrete realizations of the Shu-Schwarzschild
phase-space distribution function
(\citealt{Shu1969,BienaymeSechaud1997,BT2008}) described in
Section~2.2 of F14, with the same parameters adopted in our analytical
calculations.  The surface density of the initial conditions obtained
in this way is approximately distributed in space as an exponential
disc of scale length $\Rd=2\Kpc$. The radial and vertical velocity
dispersion on the disc plane vary approximately as
$\sigma_R\approx\sigma_{R,0}\exp\paresq{-\pare{R-R_0}/\Rs}$,
$\sigma_z\approx\sigma_{z,0}\exp\paresq{-\pare{R-R_0}/\Rs}$, where
$(\sigma_{R,0},\sigma_{z,0})=(35,15)\kmsec$, and $\Rs=5\Rd$
(\citealt{BienaymeSechaud1997}, F14).

We then integrate forward our initial conditions for a total time
$T=9\Gyr$, from $\ti=-3\Gyr$ to $\te=6\Gyr$. For $t<0$ the bar is
absent and the axisymmetric gravitational field is only given by
Model~I of \cite{BT2008}. This period of time allows the initial
conditions to become well mixed in the background
potential\footnote{The Shu-Schwarzschild distribution function is
  built on approximate integrals of motion for an axisymmetric galaxy:
  in particular, the vertical energy is a good approximate integral
  only very close to the plane. Because of Jeans' theorem
  (\citealt{BT2008}), the worse the approximation of the integrals,
  the faster the distribution function will evolve in time.}. After
the initial $3\Gyr$, we obtain a stable particle configuration, with
velocity dispersions at $(R,z)=(R_0,0)$:
$(\sigma_R,\sigma_\phi,\sigma_z)\approx(37,27,13)\kmsec$. The vertical
restoring force determines the $z$ density profile, which, at $R=R_0$
is nicely described by an exponential or $\sech^2$ profile (see
\App{app:sech}), both with scale height $h_z\approx0.3\Kpc$ close to
the plane.

The bar is introduced smoothly in the simulation at $t=0$ with the
potential defined in \Eq{pottot}, \Eq{eq:phi1fourier} and \Eq{phia},
its amplitude reaching $\epsilon=0.01$ only at $t=3\Gyr$. For $0\leq
t<3\Gyr$ the amplitude grows with time by a factor
(\citealt{Dehnen2000})
\begin{equation}
  \eta(t)=\left(\frac{3}{16}\xi^5-
  \frac{5}{8}\xi^3+\frac{15}{16}\xi+\frac{1}{2}\right),\quad
  \xi\equiv2\frac{t}{3\Gyr}-1.
\end{equation}
The amplitude then stays constant at $\epsilon=0.01$ for another
$3\Gyr$. The response becomes almost stable in the rotating frame of
the bar\footnote{Some transient effects are still present in the
  kinematics at the end of the simulation, resulting in minor
  asymmetries in the maps of the mean motions. To reach complete
  stability one should integrate for many more dynamical times, which
  would be unphysical (see, e.g., \citealt{Muhlbauer2003}).}, and the end
of the simulation can be used to compare with our analytical
predictions.

\subsubsection{Comparison with the analytical model}
\begin{figure}
  \centering
  \includegraphics[width=\columnwidth]{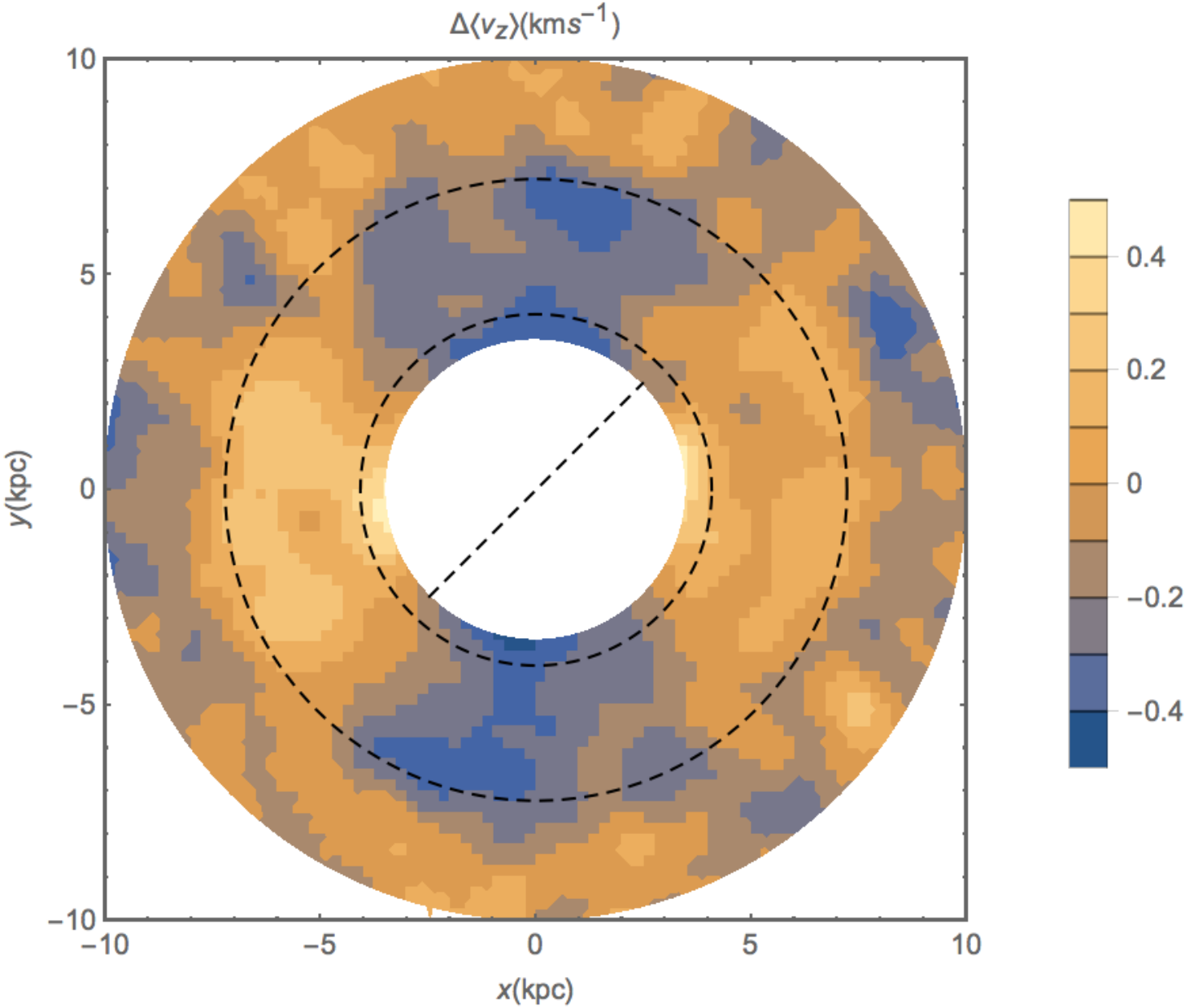}
  \includegraphics[width=\columnwidth]{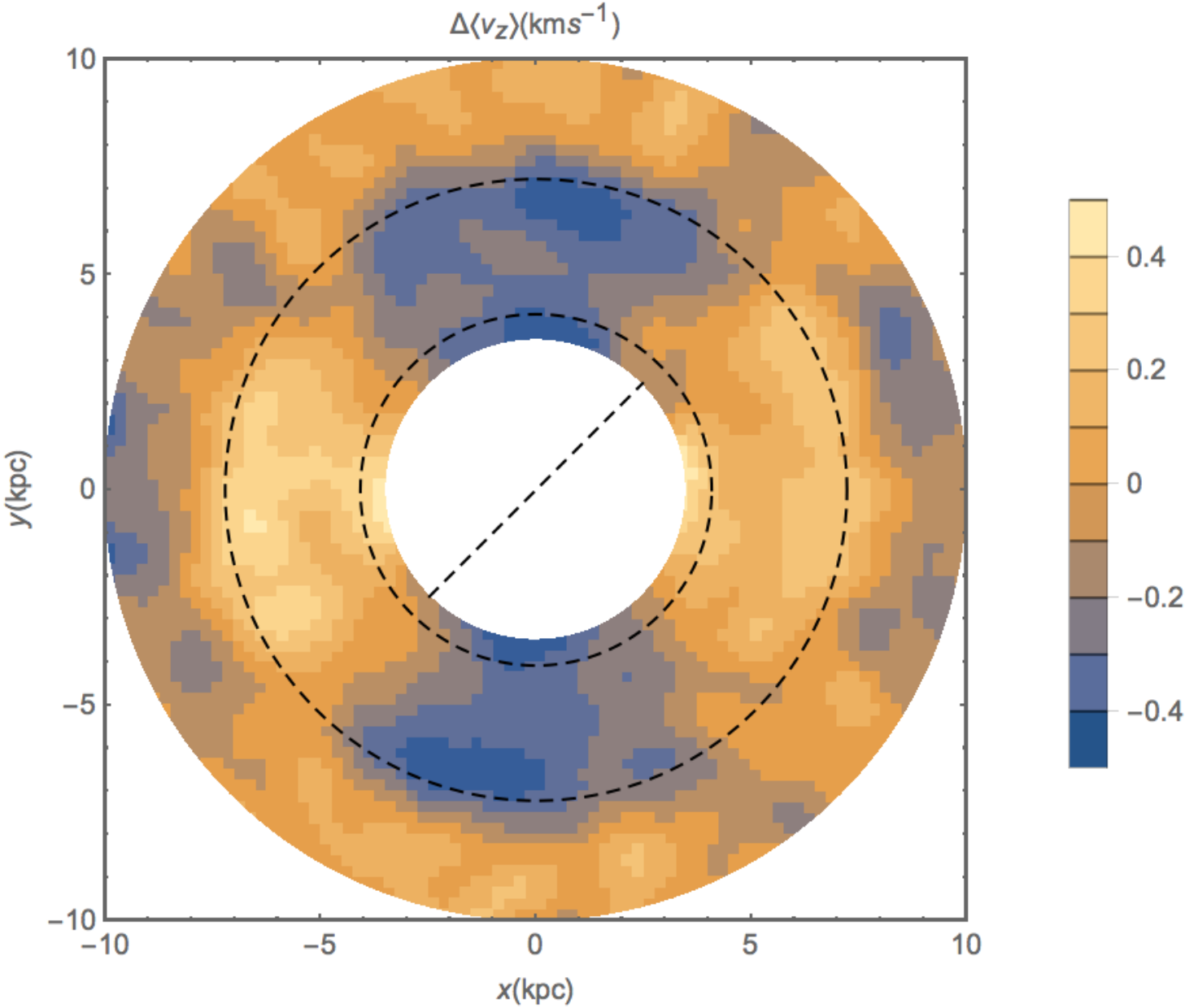}
  \caption{Difference between the average vertical motion of the
    northern and southern hemisphere of the simulated Milky Way
    $\dvz$, as a function of the $x$ and $y$ positions. The dashed
    line corresponds to the long axis of the bar, and the dashed
    circles to the position of the corotation and OLR. Top panel:
    particles at $|z|<0.3\Kpc$ only, to be directly compared with the
    analytical model. Bottom panel: all particles.}
  \label{fig:diffvz}
\end{figure}
In \Fig{fig:diffvz} we present the mean vertical kinematics as a function
of the cartesian $(x,y)$ position in the Galaxy, at the end of the
simulation. We plot the difference $\dvz\equiv\vzp - \vzn$, where
$\vzp$ ($\vzn$) is the average $v_z$ of particles found at $z>0$
($z<0$) at the end of the simulation. We present it both for particles
at $|z|<0.3\Kpc$ (top panel) and at every $z$ (bottom panel). The
averages are computed inside bins of $0.25\Kpc \times  0.25\Kpc$, also
applying a Gaussian smoothing on a scale $1\Kpc$.

\Fig{fig:diffvz} shows a very good agreement with \Fig{fig:diffvz_th},
with clear trends in the vertical kinematics, depending on the angle
from the long axis of the bar $\phi$ and on the distance from the
resonances in the same way as on \Fig{fig:diffvz_th}. The quantity
$\dvz$ appears always to be periodic with $\phi$, fluctuating twice
for $\phi \in \paresq{0,2\Pii}$, so that
$\dvz\pare{\phi}\approx\dvz\pare{\phi+\Pii}$ and
$\dvz\pare{\phi}\approx-\dvz\pare{\phi+\Pii/2}$. Notice how the phase
of the periodic oscillations changes inside and outside the OLR: the
maxima of $\dvz$ are at $\phi \sim \Pii/4\pm \Pii/2$ inside the
resonance and at $\phi \sim-\Pii/4\pm \Pii/2$ outside, as predicted by
our analytical model. 

The agreement between the analytical and numerical model is not only
qualitative, but also quantitative.
\begin{figure}
  \centering
  \includegraphics[width=\columnwidth]{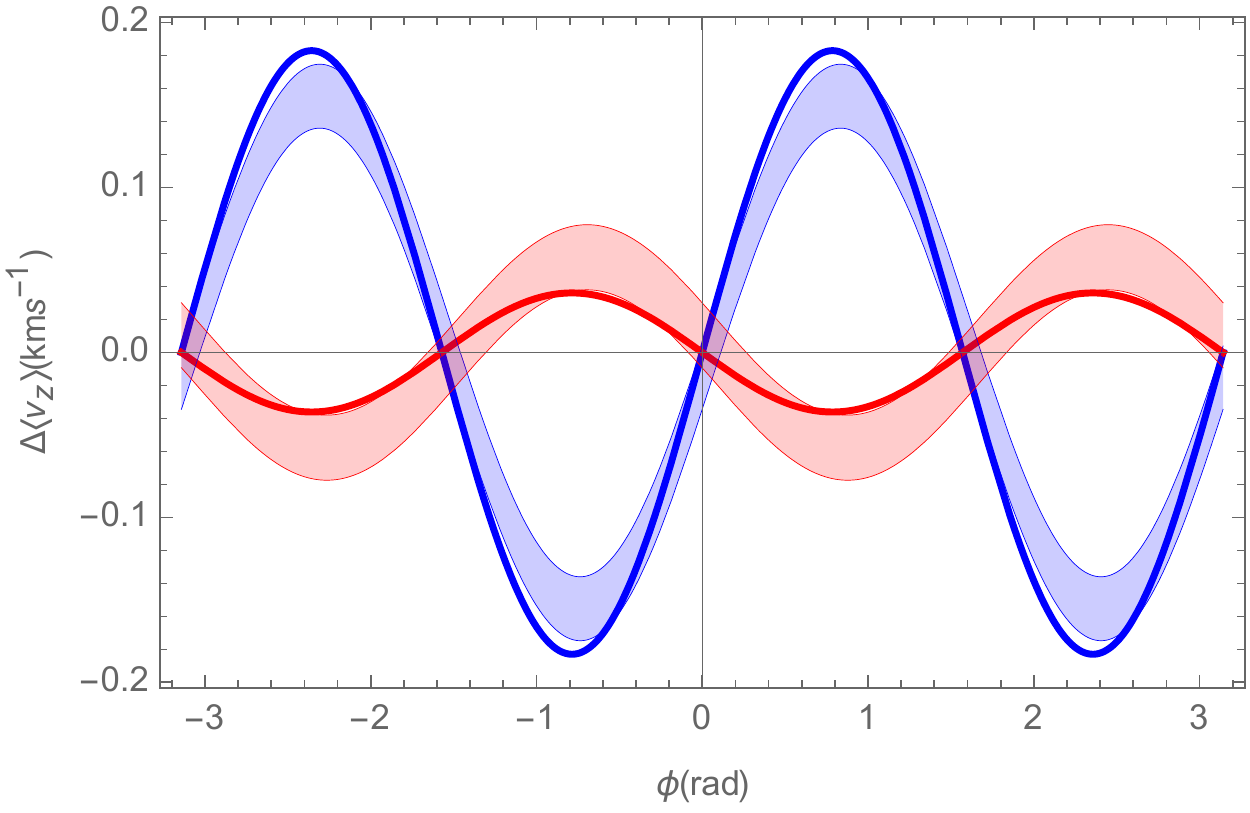}
  \caption{Predictions of the analytical model for
    $\duz(R,\phi,z)=\duz(5\Kpc,\phi,0.3\Kpc)$ (blue line) and
    $\duz(R,\phi,z)=\duz(8.5\Kpc,\phi,0.3\Kpc)$ (red line), where
    $\phi$ is measured from the long axis of the bar. The $90\%$
    confidence bands are obtained fitting $\dvz$ for particles in the
    simulation with $|R-5\Kpc|<0.2\Kpc$ (blue band) and
    $|R-8.5\Kpc|<0.2\Kpc$ (red band) with the model $a\sin(2\phi+b)$.}
  \label{fig:comp}
\end{figure}
In \Fig{fig:comp} we show the predictions of the analytical model for
$\duz(R,\phi,z)=\duz(5\Kpc,\phi,0.3\Kpc)$ (blue line) and
$\duz(R,\phi,z)=\duz(8.5\Kpc,\phi,0.3\Kpc)$ (red line), where $\phi$
is measured from the long axis of the bar. In the same plot we show
the $90\%$ confidence bands obtained fitting the model
$a\sin(2\phi+b)$ to $\dvz$ for particles in the simulation with
$|R-5\Kpc|<0.2\Kpc$ (blue band), and $|R-8.5\Kpc|<0.2\Kpc$ (red
band). The concordance between analytical model and simulations shows
how our assumptions in the analytical model are convincing. Notice how
a small phase shift is present between the analytical predictions and
the simulation. This is due to the fact that the kinematics in the
simulation did not reach a complete stability w.r.t.
the bar disturbance, as previously mentioned.

\section{Discussion and conclusions}\label{sect:concl}
In this paper we demonstrated analytically how the mean vertical
motions of stars in disc galaxies are affected by small,
non-axisymmetric perturbations of the potential in the form of Fourier
modes (e.g., bar or spiral arms). This was previously shown
numerically for spiral arms in F14, and supported by an analytical
toy-model for a pressureless fluid. The present analytical treatment
goes beyond this toy-model, and is valid for the response of realistic
disc populations to any non-axisymmetric perturbation described by
small amplitude Fourier modes.

Our general analytical solution shows that, depending on the distance
from the Lindblad resonances and azimuth in the frame of the
perturber, the mean vertical motion points towards or away from the
galactic plane, with a bulk velocity depending on the position
projected in the galactic plane and on the distance from the galactic
plane.

Although M14 did not find hints of significant vertical bulk motions
in a Galactic bar simulation, our present analytical treatment
indicates that non-zero vertical motions should be induced by the bar
too. We thus explicitly estimate them analytically, and then check for
their presence in a numerical simulation similar to that of M14. These
vertical mean motions induced by the Galactic bar are indeed modest in
magnitude, especially if compared with those observed recently in the
Solar neighborhood
(\citealt{Widrow2012},\citealt{Williams2013,Carlin2013}), but
they are well existing, as we have shown here for the very first
time. And the results of our test-particle simulation are well in line
with our analytical prediction away from the main resonances. However,
it is clear that perturbations exerting vertical forces that change
more rapidly than that of the bar with the distance from the galactic
plane (e.g., the spiral arms) do create much more significant mean
vertical velocities.

Our analytical treatment being valid close to the plane for {\it all}
the non-axisymmetric perturbations of the disc that can be described
by small-amplitude Fourier modes, it will be extremely useful for
interpreting the outcome of simulations including any such
perturbation, and to estimate how much of the observed breathing modes
can be explained by such non-axisymmetries alone. Further work should
study how the coupling of multiple internal perturbers (bar+ multiple
spirals) and external perturbers (satellite interactions themselves
exciting spiral waves, but also bending modes) is affecting the
present analytical results.

\section*{acknowledgements}
We thank the anonymous referee for useful comments. We are grateful to
James Binney for pointing out a mistake in the original version of the
manuscript. This work has been supported by a postdoctoral grant from
the {\it Centre National d'Etudes Spatiales} (CNES) for GM.

\bibliographystyle{mn2e}
\bibliography{mn-jour,bar_vzbib}

\begin{thebibliography}{37}
\expandafter\ifx\csname natexlab\endcsname\relax\def\natexlab#1{#1}\fi

\bibitem[{{Antoja} {et~al}\mbox{.}(2008){Antoja}, {Figueras}, {Fern{\'a}ndez},
  \& {Torra}}]{Antoja}
{Antoja} T., {Figueras} F., {Fern{\'a}ndez} D., {Torra} J., 2008, \aap, 490,
  135

\bibitem[{{Antoja} {et~al}\mbox{.}(2014){Antoja}, {Helmi}, {Dehnen},
  {Bienaym{\'e}}, {Bland-Hawthorn}, {Famaey}, {Freeman}, {Gibson}, {Gilmore},
  {Grebel}, {Kordopatis}, {Kunder}, {Minchev}, {Munari}, {Navarro}, {Parker},
  {Reid}, {Seabroke}, {Siebert}, {Steinmetz}, {Watson}, {Wyse}, \&
  {Zwitter}}]{Antoja2014}
{Antoja} T. {et~al.}, 2014, \aap, 563, A60

\bibitem[{{Bienaym{\'e}} {et~al}\mbox{.}(2014){Bienaym{\'e}}, {Famaey},
  {Siebert}, {Freeman}, {Gibson}, {Gilmore}, {Grebel}, {Bland-Hawthorn},
  {Kordopatis}, {Munari}, {Navarro}, {Parker}, {Reid}, {Seabroke}, {Siviero},
  {Steinmetz}, {Watson}, {Wyse}, \& {Zwitter}}]{Bienayme14}
{Bienaym{\'e}} O. {et~al.}, 2014, \aap, 571, A92

\bibitem[{{Bienayme} \& {Sechaud}(1997)}]{BienaymeSechaud1997}
{Bienayme} O., {Sechaud} N., 1997, \aap, 323, 781

\bibitem[{{Binney} \& {Tremaine}(2008)}]{BT2008}
{Binney} J., {Tremaine} S., 2008, {Galactic Dynamics: Second Edition}, {Binney,
  J.~\& Tremaine, S.}, ed. Princeton University Press

\bibitem[{{Bovy} {et~al}\mbox{.}(2015){Bovy}, {Bird}, {Garc{\'{\i}}a
  P{\'e}rez}, {Majewski}, {Nidever}, \& {Zasowski}}]{Bovy2015}
{Bovy} J., {Bird} J.~C., {Garc{\'{\i}}a P{\'e}rez} A.~E., {Majewski} S.~R.,
  {Nidever} D.~L., {Zasowski} G., 2015, \apj, 800, 83

\bibitem[{{Carlin} {et~al}\mbox{.}(2013){Carlin}, {DeLaunay}, {Newberg},
  {Deng}, {Gole}, {Grabowski}, {Jin}, {Liu}, {Liu}, {Luo}, {Yuan}, {Zhang},
  {Zhao}, \& {Zhao}}]{Carlin2013}
{Carlin} J.~L. {et~al.}, 2013, \apjl, 777, L5

\bibitem[{{Chereul} {et~al}\mbox{.}(1999){Chereul}, {Cr{\'e}z{\'e}}, \&
  {Bienaym{\'e}}}]{Chereul}
{Chereul} E., {Cr{\'e}z{\'e}} M., {Bienaym{\'e}} O., 1999, \aaps, 135, 5

\bibitem[{{Cox} \& {G{\'o}mez}(2002)}]{Cox}
{Cox} D.~P., {G{\'o}mez} G.~C., 2002, \apjs, 142, 261

\bibitem[{{Creze} {et~al}\mbox{.}(1998){Creze}, {Chereul}, {Bienayme}, \&
  {Pichon}}]{Creze98}
{Creze} M., {Chereul} E., {Bienayme} O., {Pichon} C., 1998, \aap, 329, 920

\bibitem[{{Debattista}(2014)}]{Debattista2014}
{Debattista} V.~P., 2014, \mnras, 443, L1

\bibitem[{{Dehnen}(1998)}]{Dehnen}
{Dehnen} W., 1998, \aj, 115, 2384

\bibitem[{{Dehnen}(2000)}]{Dehnen2000}
{Dehnen} W., 2000, \aj, 119, 800

\bibitem[{{Dwek} {et~al}\mbox{.}(1995){Dwek}, {Arendt}, {Hauser}, {Kelsall},
  {Lisse}, {Moseley}, {Silverberg}, {Sodroski}, \& {Weiland}}]{Dwek1995}
{Dwek} E. {et~al.}, 1995, \apj, 445, 716

\bibitem[{{Famaey} {et~al}\mbox{.}(2005){Famaey}, {Jorissen}, {Luri}, {Mayor},
  {Udry}, {Dejonghe}, \& {Turon}}]{Famaey}
{Famaey} B., {Jorissen} A., {Luri} X., {Mayor} M., {Udry} S., {Dejonghe} H.,
  {Turon} C., 2005, \aap, 430, 165

\bibitem[{{Faure} {et~al}\mbox{.}(2014){Faure}, {Siebert}, \&
  {Famaey}}]{Faure2014}
{Faure} C., {Siebert} A., {Famaey} B., 2014, \mnras, 440, 2564, (F14)

\bibitem[{{Feldmann} \& {Spolyar}(2015)}]{Feldmann}
{Feldmann} R., {Spolyar} D., 2015, \mnras, 446, 1000

\bibitem[{{G{\'o}mez} {et~al}\mbox{.}(2013){G{\'o}mez}, {Minchev}, {O'Shea},
  {Beers}, {Bullock}, \& {Purcell}}]{Gomez2013}
{G{\'o}mez} F.~A., {Minchev} I., {O'Shea} B.~W., {Beers} T.~C., {Bullock}
  J.~S., {Purcell} C.~W., 2013, \mnras, 429, 159

\bibitem[{{Kuijken} \& {Gilmore}(1991)}]{KuijkenGil91}
{Kuijken} K., {Gilmore} G., 1991, \apjl, 367, L9

\bibitem[{{Kuijken} \& {Tremaine}(1991)}]{Kuijken1991}
{Kuijken} K., {Tremaine} S., 1991, in Dynamics of Disc Galaxies, {Sundelius}
  B., ed., p.~71

\bibitem[{{Lin} \& {Shu}(1964)}]{LinShu}
{Lin} C.~C., {Shu} F.~H., 1964, \apj, 140, 646

\bibitem[{{Monari} {et~al}\mbox{.}(2014){Monari}, {Helmi}, {Antoja}, \&
  {Steinmetz}}]{Monari2014}
{Monari} G., {Helmi} A., {Antoja} T., {Steinmetz} M., 2014, \aap, 569, A69, (M14)

\bibitem[{{M{\"u}hlbauer} \& {Dehnen}(2003)}]{Muhlbauer2003}
{M{\"u}hlbauer} G., {Dehnen} W., 2003, \aap, 401, 975

\bibitem[{{Patsis} \& {Grosbol}(1996)}]{Patsis}
{Patsis} P.~A., {Grosbol} P., 1996, \aap, 315, 371

\bibitem[{{Read}(2014)}]{Read2014}
{Read} J.~I., 2014, Journal of Physics G Nuclear Physics, 41, 063101

\bibitem[{{Shu}(1969)}]{Shu1969}
{Shu} F.~H., 1969, \apj, 158, 505

\bibitem[{{Siebert} {et~al}\mbox{.}(2003){Siebert}, {Bienaym{\'e}}, \&
  {Soubiran}}]{Siebert03}
{Siebert} A., {Bienaym{\'e}} O., {Soubiran} C., 2003, \aap, 399, 531

\bibitem[{{Siebert} {et~al}\mbox{.}(2012){Siebert}, {Famaey}, {Binney},
  {Burnett}, {Faure}, {Minchev}, {Williams}, {Bienaym{\'e}}, {Bland-Hawthorn},
  {Boeche}, {Gibson}, {Grebel}, {Helmi}, {Just}, {Munari}, {Navarro}, {Parker},
  {Reid}, {Seabroke}, {Siviero}, {Steinmetz}, \& {Zwitter}}]{Siebert2012}
{Siebert} A. {et~al.}, 2012, \mnras, 425, 2335

\bibitem[{{Siebert} {et~al}\mbox{.}(2011){Siebert}, {Famaey}, {Minchev},
  {Seabroke}, {Binney}, {Burnett}, {Freeman}, {Williams}, {Bienaym{\'e}},
  {Bland-Hawthorn}, {Campbell}, {Fulbright}, {Gibson}, {Gilmore}, {Grebel},
  {Helmi}, {Munari}, {Navarro}, {Parker}, {Reid}, {Siviero}, {Steinmetz},
  {Watson}, {Wyse}, \& {Zwitter}}]{Siebert2011}
{Siebert} A. {et~al.}, 2011, \mnras, 412, 2026

\bibitem[{{Toomre}(1964)}]{Toomre}
{Toomre} A., 1964, \apj, 139, 1217

\bibitem[{{Weinberg}(1994)}]{Weinberg1994}
{Weinberg} M.~D., 1994, \apj, 420, 597

\bibitem[{{Widrow} {et~al}\mbox{.}(2014){Widrow}, {Barber}, {Chequers}, \&
  {Cheng}}]{Widrow2014}
{Widrow} L.~M., {Barber} J., {Chequers} M.~H., {Cheng} E., 2014, \mnras, 440,
  1971

\bibitem[{{Widrow} \& {Bonner}(2015)}]{Widrow2015}
{Widrow} L.~M., {Bonner} G., 2015, \mnras, 450, 266

\bibitem[{{Widrow} {et~al}\mbox{.}(2012){Widrow}, {Gardner}, {Yanny},
  {Dodelson}, \& {Chen}}]{Widrow2012}
{Widrow} L.~M., {Gardner} S., {Yanny} B., {Dodelson} S., {Chen} H.-Y., 2012,
  \apjl, 750, L41

\bibitem[{{Williams} {et~al}\mbox{.}(2013){Williams}, {Steinmetz}, {Binney},
  {Siebert}, {Enke}, {Famaey}, {Minchev}, {de Jong}, {Boeche}, {Freeman},
  {Bienaym{\'e}}, {Bland-Hawthorn}, {Gibson}, {Gilmore}, {Grebel}, {Helmi},
  {Kordopatis}, {Munari}, {Navarro}, {Parker}, {Reid}, {Seabroke}, {Sharma},
  {Siviero}, {Watson}, {Wyse}, \& {Zwitter}}]{Williams2013}
{Williams} M.~E.~K. {et~al.}, 2013, \mnras, 436, 101

\bibitem[{{Xu} {et~al}\mbox{.}(2015){Xu}, {Newberg}, {Carlin}, {Liu}, {Deng},
  {Li}, {Sch{\"o}nrich}, \& {Yanny}}]{Xu}
{Xu} Y., {Newberg} H.~J., {Carlin} J.~L., {Liu} C., {Deng} L., {Li} J.,
  {Sch{\"o}nrich} R., {Yanny} B., 2015, \apj, 801, 105

\bibitem[{{Yanny} \& {Gardner}(2013)}]{YannyGardner2013}
{Yanny} B., {Gardner} S., 2013, \apj, 777, 91

\end{thebibliography}

\begin{appendix}

\section{Results for a sech$^2$ vertical density distribution}\label{app:sech}
\begin{figure}
  \centering
  \includegraphics[width=\columnwidth]{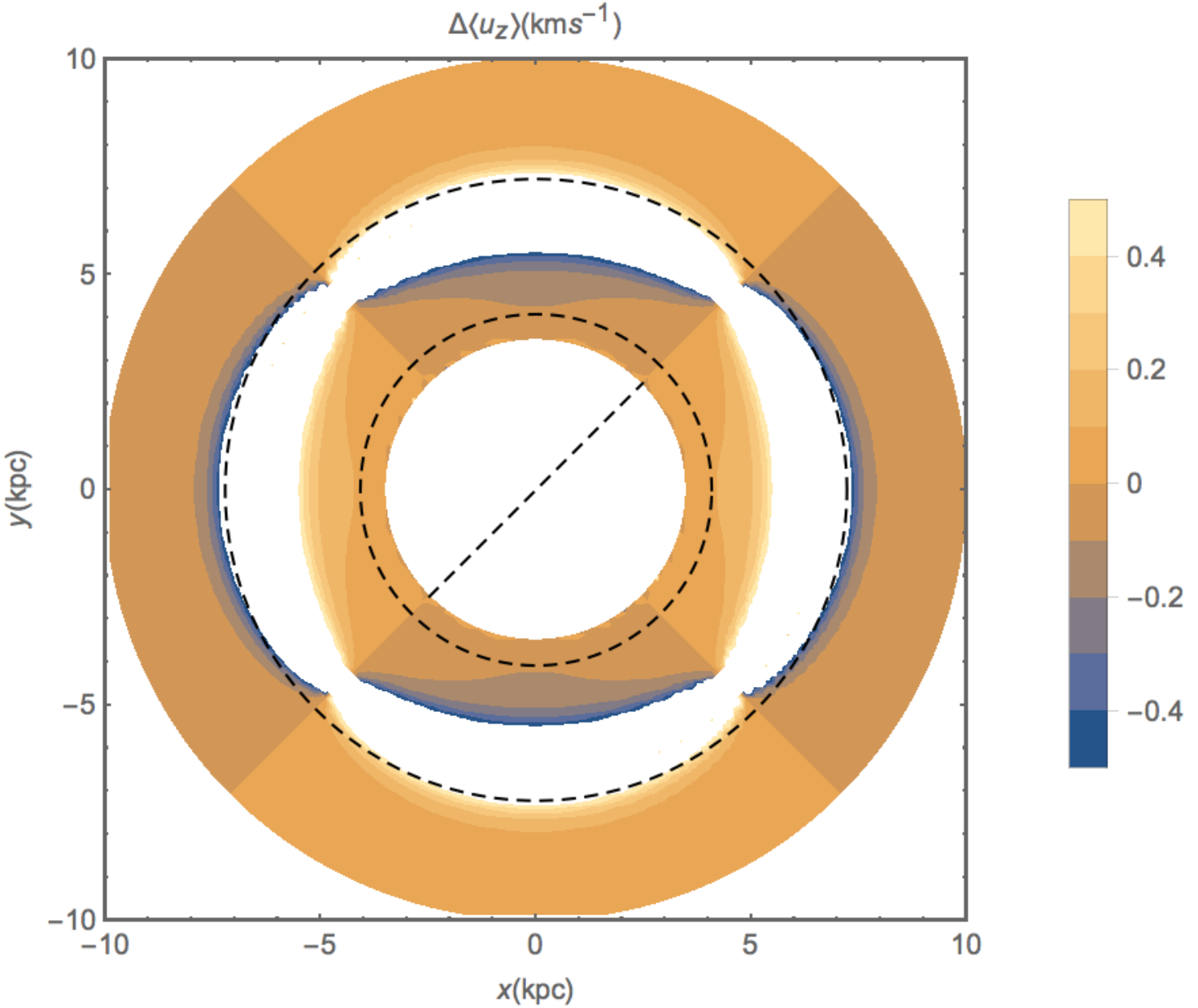}
  \caption{As in \Fig{fig:diffvz_th}, but for a density distribution
    depending on $z$ as $\sech^2(z/h_z)$.}
  \label{fig:diffvzsech_th}
\end{figure}

\begin{figure}
  \centering
  \includegraphics[width=\columnwidth]{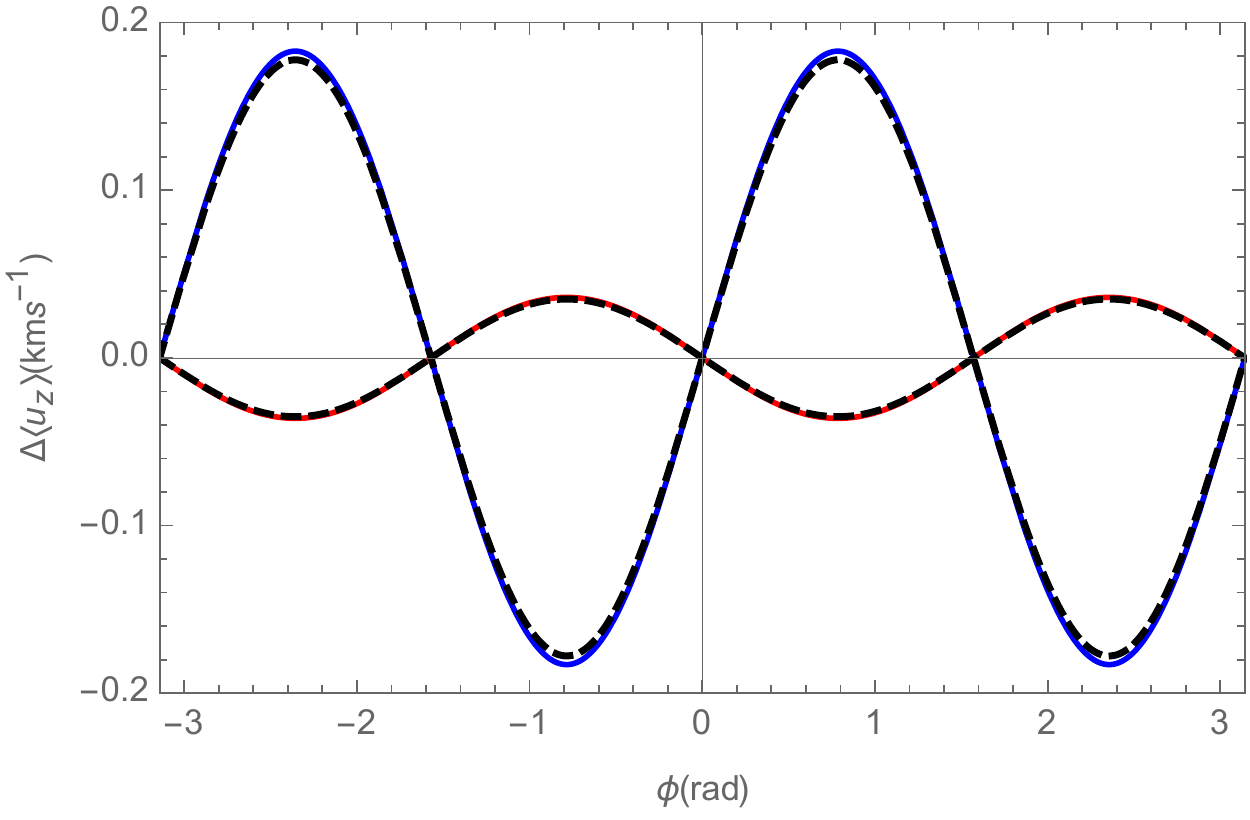}
  \caption{Comparison between the analytical $\duz$ in the case of a
    density distribution depending on $z$ as $\exp\pare{-|z|/h_z}$
    (colored lines) and $\sech^2(z/h_z)$ (black dashed lines). The
    potential parameters and scale height are as in
    \Fig{fig:comp}. The blue line correspond to $R=5\Kpc$, the red
    line to $R=8.5\Kpc$.}
  \label{fig:compsech}
\end{figure}

We report here the analytical results for $u_z$ and $\duz$ obtained
with a different unperturbed vertical density distribution, namely
\begin{equation}
  \rho_0\pare{R,z}=\rho_0\pare{0,0}\exp\pare{-\frac{R}{\Rd}}\sech^2\pare{\frac{z}{h_z}}.
\end{equation}
With this continuously differentiable choice of density, and following the same steps as in
\Sec{sect:dynamics}, \Eqs{eq:uz}{eq:uzint} become
\begin{align}\label{eq:uzsech}
  u_z(z) &=-\epsilon h_z\cosh^2\pare{\frac{z}{h_z}}\biggl\{
    \calG(0)\tanh\pare{\frac{z}{h_z}} \nonumber \displaybreak[0] \\
    &+\ddp{^2\calG}{z^2}(0)\biggl[\frac{h_z^2\Pii^2}{24}+\frac{h_z^2}{2}\Li_2\pare{-\eexp^{-2z/h_z}}
      \nonumber \displaybreak[0] \\
    &-h_z\ln\pare{1+\eexp^{-2z/h_z}}z+\frac{1}{2}\pare{\tanh\pare{\frac{z}{h_z}}-1}z^2\biggr]\biggr\},
\end{align}
and,
\begin{align}\label{eq:uzintsech}
  \duz &=-2\epsilon \coth\pare{\frac{z}{h_z}}\biggl\{
  -h_z\calG(0)\ln\paresq{\cosh\pare{\frac{z}{h_z}}} \nonumber \\
  &+\ddp{^2\calG}{z^2}(0)\biggl[-\frac{h_z^3}{4}
   \Li_3\left(-\eexp^{-\frac{2 z}{h_z}}\right)-\frac{h_z^3}{2}
   \Li_3\left(-\eexp^{\frac{2 z}{h_z}}\right) \nonumber \\
   &-\frac{9h_z^3}{16}\zeta(3)-\frac{h_z^2}{2}\Li_2\left(-\eexp^{-\frac{2 z}{h_z}}\right)z+\frac{h_z^2}{2}
   \Li_2\left(-\eexp^{\frac{2 z}{h_z}}\right)z \nonumber \\
   &-\frac{\Pii^2h_z^2}{24}z+\frac{h_z}{2}\log\left(\eexp^{-\frac{2 z}{h_z}}+1\right)z^2
   +\frac{z^3}{3}\biggr]\biggr\},
\end{align}
where the ``polylogarithm'' function $\Li_s(z)$ is defined by the power series
\begin{equation}
  \Li_s(z)=\sum_{k=1}^{\infty}\frac{z^k}{k^s},
\end{equation}
and the Riemann $\zeta(s)$ function by 
\begin{equation}
  \zeta(s)=\sum_{k=1}^{\infty}k^{-s}.
\end{equation}
Using these new formulae \Eqs{eq:uzsech}{eq:uzintsech}, we obtain
\Fig{fig:diffvzsech_th}, corresponding to \Fig{fig:diffvz_th}, and
\Fig{fig:compsech}, corresponding to \Fig{fig:comp}. These figures
show how in the bar case the relative difference between the predicted
bulk motions for the case of a $\sech^2(z/h_z)$, and
$\exp\pare{-|z|/h_z}$ density distributions are tiny (of the order of
$5 \times 10^{-3}\kmsec$ at most). We note that the initial
axisymmetric stellar population in our test particle simulation is
well fit close to the plane by $\sech^2$ and exponential distributions
with the same scale height $h_z=0.3\Kpc$. However, we note that the
$\sech^2$ profile is actually a better fit, as expected from the form
of the Shu-Schwarzschild distribution function used for the initial
conditions.

\section{Estimating $\uphiz$ near the plane}\label{app:uphiz}
Jeans' equation for $u_\phi$, in the case of an axisymmetric system,
and neglecting mixed terms can be rewritten as (\citealt{BT2008})
\begin{equation}
  \ddp{\pare{\rho_0\sigma_{R,0}^2}}{R}+ 
  \rho_0\pare{\frac{\sigma_{R,0}^2-\sigma_{\phi,0}^2-\uphiz^2}{R}+\ddp{\Phi_0}{R}}=0.
\end{equation}
Differentiating twice with the respect of $z$, assuming an
exponential disc, and that the velocity dispersion near the plane is
approximately constant with $z$, we obtain
\begin{equation}
  \uphiz^2(R,z)+\uphiz(R,z)\ddp{^2\uphiz}{z^2}(R,z)=\frac{R}{2}\ddp{^3\Phi_0}{R\partial z^2}(R,z),
\end{equation}
which, estimated at $z=0$, is
\begin{equation}
  \ddp{^2\uphiz}{z^2}(R,0)=\frac{R}{2\uphiz(R,0)}\dd{\nu^2}{R}(R),
\end{equation}
and 
\begin{equation}
  \nu^2(R)\equiv\ddp{^2\Phi_0}{z^2}(R,0),
\end{equation}
is the square of the vertical frequency.

\end{appendix}

\label{lastpage}

\end{document}